# Isotopic dependence of impurity transport driven by ion temperature gradient turbulence

Weixin Guo, Lu Wang [a)] and Ge Zhuang

*State Key Laboratory of Advanced Electromagnetic Engineering and Technology, School of Electrical and Electronic Engineering, Huazhong University of Science and Technology, Wuhan, Hubei 430074, China*

[a)] E-mail: luwang@hust.edu.cn

Hydrogenic ion mass effects, namely the isotopic effects on impurity transport driven by ion temperature gradient (ITG) turbulence are investigated using gyrokinetic theory. For non-trace impurities, changing from hydrogen (H) to deuterium (D), and to tritium (T) plasmas, the outward flux for lower (higher) ionized impurities or for lighter (heavier) impurities is found to decrease (increase), although isotopic dependence of ITG linear growth rate is weak. This is mainly due to the decrease of outward (inward) convection, while the isotopic dependence of diffusion is relatively weak. In addition, the isotopic effects reduce (enhance) the impurity flux of fully ionized carbon ($C^{6+}$) for weaker (stronger) magnetic shear. In trace impurity limit, the isotopic effects are found to reduce the accumulation of high-$Z$ tungsten (W). Moreover, the isotopic effects on the peaking factor (PF) of trace high-$Z$ W get stronger with stronger magnetic shear.

## I. Introduction

In fusion plasmas, the non-hydrogenic ions, i.e., impurities produced mainly from the plasma facing materials interaction are inevitable elements. Especially, helium ash generated by deuterium (D)-tritium (T) reaction in fusion reactor is another species of intrinsic impurity. Sometimes, deliberate impurity seeding is also an important impurity source. Although impurities create a radiation belt in the edge for exhausting continuous heat, accumulation of them, especially high-$Z$ impurity such as tungsten (W) in the core can cause the fuel dilution and power loss through radiation, leading to plasma density, temperature and fusion power degradation, [1-3] which is disastrous for the success of fusion reactor. Here, $Z$ is the charge number of impurity. Accordingly, comprehensive understanding of physics mechanisms of impurity transport, particularly the conditions of avoiding the impurity accumulation in the core is of great significance for fusion plasmas.

Actually, impurity transport is a universal issue in magnetic confinement plasmas. Apart from the helical systems[4] and reversed field pinch (RFP),[5] impurity transport has also been intensively studied in tokamak plasmas by neo-classical theory[6-8] and anomalous transport theory.[9-34] The anomalous transport is attributed to the presence of drift wave-type turbulence. For turbulent impurity transport, the theoretical and simulation works[10-21] all hitherto existing mainly focused on ion temperature gradient mode (ITG) and trapped electron mode (TEM) turbulence.

In the published literatures of turbulent impurity transport, the most used simplification is trace (tested, passive) impurity approximation[13-16, 30-32] or dilution model[11] especially for heavy metal impurities.[35] But, impurities can have non-negligible influence on turbulence.[36-45] For example, the effects of impurity concentration on the linear growth rate depend on impurity density gradient,[37, 44, 45] but the critical value is not quantitatively pointed out. It further motivates impurity transport study by considering the effects of impurity profile on turbulence called non-trace impurity transport.[10-12, 30, 38, 46] No matter for non-trace or trace impurity transport, various physical interpretations regarding to the magnitude or the sign of the impurity transport as well as its parametric dependencies, such as $Z$, impurity mass number ($A_z$)[11, 14, 18] and magnetic shear[30, 32, 47] have been widely studied.

The isotopic effects are also longstanding issues. It is found that confinement of momentum, particle and energy is improved as changing from hydrogen (H) to D or D-T plasmas. In different operation regimes, the mass scaling is found to be widely varied in the range $A_i^{0-0.85}$,[48] where $A_i$ is the mass number of main ions. Particularly, the isotopic scaling of the energy confinement time, $A_i^{0.6}$ was found for ITG turbulence.[49] The ITG turbulence is thought to be the dominant candidate for anomalous transport in the bulk plasma region. Recently, isotopic effects on impurity transport are also experimentally investigated in both stellarator and tokamaks.[50, 51] Impurity confinement is found to be decreased by increasing the D/H ratio in TJ-II.[50] In JET,[51]

high-$Z$ impurity transport level from boundary to core may be higher with higher effective charge number and higher effective mass number. The isotopic effects on light or medium mass impurities, to our knowledge, have not been investigated in tokamaks yet. Theoretical and simulation studies of isotopic effects on micro-turbulence [52] has been reported. Especially, effective mass dependence of linear growth rate of ITG and TEM were also studied by including impurities. [53, 54] However, theoretical study of isotopic effects on impurity transport is still lacking.

The aim of this work is to explore isotopic effects on impurity transport driven by ITG turbulence. In this paper, the effects of changing hydrogen isotopes from lighter to heavier are referred as to isotopic effects. We focus on the quasi-linear (QL) turbulent impurity transport in sheared slab geometry without the assumption of trace impurities. The impurity flux is divided into diffusive and convective parts expressed as $\Gamma_z = -D_z \nabla n_{0z} + V_z n_{0z}$, where $\Gamma_z$ represents the impurity flux, $D_z$ and $V_z$ are the diffusivity and convective velocity, respectively. Various parameters such as $A_z$ and $Z$ as well as magnetic shear parameter ($S$) scans for isotopic effects on impurity flux and the corresponding diffusivity and convective velocity are investigated. The tolerance for heavy impurities such as W in the core for the stable plasma operation is about 0.01%. [35] For this concentration, our results can be reduced to trace impurity limit. Then, the isotopic effects on impurity peaking factor (PF) defined as $-RV_z/D_z$ are accordingly studied. H, D and T are chosen as the main isotope working gases. The main results in the present work are summarized in Table. I.

Table I. Overview of results: parametric scans for isotopic effects on normalized impurity flux $\hat{\Gamma}_z$, convective velocity $\hat{V}_z$, diffusivity $\hat{D}_z$, PF, respectively.

| Parameters | Non-trace impurity | | Trace impurity |
|---|---|---|---|
| | $\hat{\Gamma}_z$ ($\hat{V}_z$) | $\hat{D}_z$ | PF |
| $Z$ | $Z$: unfavorable[a] <br> $Z>5$: favorable[b] | | $Z<6$: unfavorable <br> $Z\geq 7$: favorable |
| $A_z$ | $A_z \leq 12$: unfavorable <br> $A_z \geq 14$: favorable | Weak[c] | ---[d] |
| $\|S\|$ | For $C^{6+}$: <br> $\|S\|<0.15$: unfavorable <br> $\|S\|\geq 0.15$: favorable | | For $W^{45+}$: <br> Favorable |

[a]Unfavorable ([b]Favorable): isotopic effects in this parameter region are unfavorable (favorable) for increase of outward flux (negative PF) or decrease of inward flux (positive PF).

[c]Weak: isotopic effects on impurity transport in this parameter region are weak.

[d]---: no scan about this parameter.

The remainder of this paper is organized as follows. In Sec. II, the dispersion equation in the presence of non-trace impurities is derived, and the isotopic effects on linear growth rate and mode frequency of ITG are given. The isotopic effects on impurity transport driven by ITG turbulence are discussed in Sec. III. Finally, a brief summary and some discussions are presented in Sec. IV.

## II. Isotopic dependence of ITG instability with non-trace impurities.

In this section, we study the isotopic dependence of ITG mode in the presence of impurities including both the dilution and the impurity density profile effects. Both the perturbed gyrocenter distribution functions for ions and impurities can be separated into the adiabatic part and non-adiabatic part, i.e., $\delta f_\alpha = -\frac{Z_\alpha e \langle \delta \phi_{gc} \rangle}{T_\alpha} \overline{F}_{\alpha 0} + \delta g_\alpha$. Here, the equilibrium distribution function $\overline{F}_{\alpha 0}$ is assumed to be Maxwellian. The subscript $\alpha = i, z$ denotes the ions and impurities, respectively, $Z_\alpha$ is the charge number ($Z_\alpha = 1$ for hydrogenic ions, $Z_\alpha = Z$ for impurities), $\delta\phi_{gc}$ is the elecrostatic potential fluctuation, $T_\alpha$ is the temperature of ions or impurities. The $\langle ... \rangle$ denotes a gyrophase average. In the following, we consider the magnetic field mainly along the *z* direction and the shear component along the *y* direction expressed as $\vec{B} = B\left(\vec{e}_z + \frac{x}{L_s}\vec{e}_y\right)$, [55] where $L_s$ is the scale length of magnetic shear ($L_s > (<) 0$ corresponds to normal (negative) magnetic shear), *x, y,* and *z* directions correspond to the radial, poloidal, and toroidal directions in tokamak configuration. The linearized gyrocenter Vlasov equation for the non-adiabatic response is given as [56]

$$-i\omega\delta g_\alpha + ik_\parallel \overline{U}\delta g_\alpha = \frac{c}{B}ik_y\langle\delta\phi_{gc}\rangle\overline{\nabla F}_{\alpha 0} - i\omega\frac{e}{T_\alpha}\overline{F}_{\alpha 0}\langle\delta\phi_{gc}\rangle, \quad (1)$$

where $\omega$ is the characteristic fluctuation frequency, $\overline{U}$ is the parallel gyrocenter velocity, $c$ is the light velocity in vacuum, $k_\parallel$ and $k_y$ are wave vectors along the parallel direction with respect to the magnetic field $\vec{B}$ and *y* direction, respectively. In the fluid regime, most of the ions and impurities satisfy the relationship $k_\parallel \overline{U} \ll \omega$, so that we expand the propagator up to the second order:

$$\frac{1}{\omega - k_\parallel \overline{U}} \approx \frac{1}{\omega} + \frac{k_\parallel \overline{U}}{\omega^2} + \frac{k_\parallel^2 \overline{U}^2}{\omega^3}. \quad (2)$$

The electrostatic potential fluctuation can be written in Fourier space $\delta\phi(\vec{r}) = \sum_k \delta\phi_k(x)e^{-i\omega t + ik_z z + ik_y y}$. For the sake of simplicity, we take $k_\parallel = k_\parallel' x$ with $k_\parallel' = \frac{k_y}{L_s}$ and *x* being the distance from the location of rational surface. After pull-back transformation and integration in phase space, we can obtain the perturbed density for main ions and impurities

$$\delta n_{\alpha,k} = -\left\{\frac{\omega_{*\alpha}}{\omega} + \left[1 - \frac{\omega_{*\alpha}}{\omega}(1+\eta_\alpha)\right]\left[b_{\alpha\perp} - \frac{1}{2}\left(\frac{k'_\parallel v_{th\alpha}}{\omega}\right)^2 x^2\right]\right\}\frac{Z_\alpha e \delta\phi_k(x)}{T_\alpha} n_{0\alpha}. \qquad (3)$$

We should point out that the lowest order of finite Larmor radius (FLR) effects and parallel compression terms are retained. Here, $\omega_{*\alpha} = -\frac{k_y}{L_{n_\alpha}}\frac{cT_\alpha}{Z_\alpha eB}$ is the diamagnetic drift frequency, $L_{n_\alpha}^{-1} = -\frac{\partial \ln n_{0\alpha}}{\partial r}$, $\eta_\alpha = \frac{L_{n_\alpha}}{L_{T_\alpha}}$ with $L_{T_\alpha}^{-1} = -\frac{\partial \ln T_\alpha}{\partial r}$, $b_{\alpha\perp} = \frac{1}{2}k_\perp^2 \rho_\alpha^2$, $k_\perp^2 = k_y^2 - \frac{\partial^2}{\partial x^2}$, $\rho_\alpha = \frac{v_{th\alpha}}{\Omega_\alpha}$ is the thermal gyroradius, $v_{th\alpha}$ is the thermal velocity, $\Omega_\alpha = \frac{Z_\alpha eB}{m_\alpha c}$ is the particle cyclotron frequency with $m_\alpha$ being the mass of ions or impurities. For adiabatic electrons i.e., $\delta n_{e,k} = \frac{e\delta\phi_k}{T_e}$ with $T_e$ being the electron temperature, the radial differential dispersion eigenmode equation can be obtained from the quasi-neutrality equation $\delta n_e = \delta n_i + Z\delta n_z$, i.e.,

$$\left(A_W \frac{\partial^2}{\partial \bar{x}^2} + B_W - C_W \bar{x}^2\right)\delta\phi_k(x) = 0, \qquad (4)$$

with coefficients $A_W = \frac{1}{2}\sum_\alpha \frac{\rho_\alpha^2}{\rho_s^2} g_\alpha \left(1 - \frac{\omega_{*\alpha}}{\omega}(1+\eta_\alpha)\right)$,

$$B_W = -\left\{1 + \sum_\alpha g_\alpha\left[\frac{\omega_{*\alpha}}{\omega} + \frac{1}{2}k_y^2\rho_\alpha^2\left(1 - \frac{\omega_{*\alpha}}{\omega}(1+\eta_\alpha)\right)\right]\right\},$$

$$C_W = -\frac{1}{2}\rho_s^2 \sum_\alpha g_\alpha \left(\frac{k'_\parallel v_{th\alpha}}{\omega}\right)^2\left(1 - \frac{\omega_{*\alpha}}{\omega}(1+\eta_\alpha)\right).$$

This is a Weber equation. Here, $\bar{x} = \frac{x}{\rho_s}$ with $\rho_s = \frac{c_s}{\Omega_i}$, $c_s$ is the ion acoustic velocity, $g_i = \tau_i(1-Zf_c)$ and $g_z = \tau_z Z^2 f_c$ are weighting factors corresponding to ions and impurities with $\tau_i = T_e/T_i$, $\tau_z = T_e/T_z$, and $f_c = n_{0z}/n_{0e}$ being the impurity concentration. Solving Eq. (4), the corresponding eigenfunction is thus given as

$$\delta\phi_k(x) \sim \exp(-i|S|\frac{F(\hat{\omega})}{\hat{\omega}}\frac{x^2}{2}), \tag{5}$$

and the dispersion equation can also be followed as

$$\hat{\omega}^2 + (b_{iy}\tau_i\tilde{f}_z)\hat{\omega}^2 + (b_{iy}\tau_i\overline{K}\tilde{f}_z)\hat{\omega} - \hat{\omega} + i(2\ell+1)|S|F(\hat{\omega})(\hat{\omega}+\overline{K})\tilde{f}_z = 0. \tag{6}$$

Here, $\ell$ is the order of the Hermite polynomial. The other symbols are in the following: $\hat{\omega}=\frac{\omega}{\omega_{*e}}$, $\omega_{*e}=\frac{k_y}{L_{n_e}}\frac{cT_e}{eB}$, $L_{n_e}^{-1}=-\frac{\partial \ln n_{0e}}{\partial r}$, $b_{iy}=\frac{1}{2}k_y^2\rho_i^2$,

$\tilde{f}_z = 1 - Zf_c + \frac{A_z}{A_i}f_c$, $K_\alpha=\frac{(1+\eta_\alpha)}{\tau_\alpha}$, $\overline{K}=\frac{(1-Zf_c)L_{ei}K_i + \frac{A_z}{A_iZ}f_cL_{ez}K_z}{\tilde{f}_z}$, $S=\frac{L_{n_e}}{L_s}$,

$L_{e\alpha}=\frac{L_{n_e}}{L_{n_\alpha}}$, $F(\hat{\omega})=\left(1-\left(1-\frac{A_i^2Z^2}{A_z^2}\right)\frac{A_z}{A_i}\frac{f_c(\hat{\omega}+L_{ez}K_z/Z)}{\tilde{f}_z(\hat{\omega}+\overline{K})}\right)$. We should note that it will be reduced to unsheared case if magnetic shear approaches to 0. This is not in the scope of this paper. The range of $|S|$ is from 0.05 to 1 in the later part of this paper. The absolute value of $S$ is required by the 'outgoing' energy condition.[57] In addition, the equilibrium quasi-neutrality equation imposes the relationship $Zf_cL_{ez}+(1-Zf_c)L_{ei}=1$.

If $T_z=T_i$ is further assumed, we have $\eta_z=\eta_i\left(\frac{1}{L_{ez}}-Zf_c\right)/(1-Zf_c)$. Ref. [46] used an approximation of $A_z/A_i=Z$ which can make $F(\hat{\omega})=1$. However, the charge number $Z$ and mass number $A_z$ of various impurities in our work are realistic. Therefore, we cannot use the same approximation as Ref. [46]. To obtain an analytic solution, we just artificially take $F(\hat{\omega})\approx 1$, which is equivalent to an approximation of cancellation of impurity contributions to $C_W$ and $A_W$. This might result in inconsistency with keeping impurity contribution in other terms, which should be carefully treated in future. Then, the dispersion equation can be reduced as

$$(1+b_{iy}\tau_i\tilde{f}_z)\hat{\omega}^2 +(b_{iy}\tau_i\overline{K}\tilde{f}_z -1+i(2\ell+1)|S|)\hat{\omega}+i(2\ell+1)|S|\overline{K}\tilde{f}_z =0 \quad (7)$$

For $2|S|[1+\overline{K}\tilde{f}_z(2+(2\tilde{f}_z-1)b_{iy}\tau_i)] > (b_{iy}\tau_i\overline{K}\tilde{f}_z-1)^2$, we can obtain the normalized real frequency and linear growth rate with $\ell=0$

$$\hat{\omega}_r = -\frac{\sqrt{|S|[1+\overline{K}\tilde{f}_z(2+(2\tilde{f}_z-1)b_{iy}\tau_i)]}}{2(1+b_{iy}\tau_i\tilde{f}_z)}, \quad (8)$$

$$\hat{\gamma} = \frac{\sqrt{|S|[1+\overline{K}\tilde{f}_z(2+(2\tilde{f}_z-1)b_{iy}\tau_i)]}-|S|}{2(1+b_{iy}\tau_i\tilde{f}_z)}, \quad (9)$$

respectively. Comparing to Ref. 55, we add the influence of impurities on mode instability and keep the FLR effects and all terms related to magnetic shear. Here, we do not consider resonant ion effects, [58] which might need to be investigated in depth in future. From Eq. (9), we can see that the dependence of linear growth rate on impurity concentration relies on impurity density profile, which mainly comes from the effective ITG drive $\overline{K}\tilde{f}_z = K_i - \frac{Zf_cL_{ez}}{\tau_i}\left[1-\frac{A_z}{A_iZ^2}+\left(1-\frac{A_z}{A_iZ^2}\frac{1}{L_{ez}}\right)\eta_i\right]$. It is obvious that the condition, $1-\frac{A_z}{A_iZ^2}+\left(1-\frac{A_z}{A_iZ^2}\frac{1}{L_{ez}}\right)\eta_i = 0$ determines a critical impurity density gradient $L_{ez,c}$ for discriminating destabilization and stabilization effects

$$L_{ez,c} = \frac{A_z}{A_iZ^2}\left[1+\left(1-\frac{A_z}{A_iZ^2}\right)/\eta_i\right]^{-1}. \quad (10)$$

For $L_{ez}>(<)L_{ez,c}$, the effective ITG drive is reduced (enhanced) by the presence of impurities. From the results of Eqs. (8) and (9), the isotopic dependence as well as the parametric dependence such as impurity concentration and impurity profile of the ITG instability are examined, which are shown in Fig.1. We take the following parameters: $|S|=0.1$, $\tau_i=\tau_z=1$ ($T_z=T_i=T_e$), $R/L_{n_e}=2$, $\eta_i=3$, $k_y\rho_s=0.2$, $f_c=0.01$, and fully ionized carbon ($C^{6+}$) is taken as the impurity unless otherwise stated. The isotopic components are H, D, T, respectively.

The normalized growth rate is shown in Fig. 1 (a), (c), and mode frequency is shown in Fig. 1(b), (d) as a function of $L_{ez}$ with different $f_c$ and different isotopic components. As shown in Fig. 1(a) and (b), both the normalized growth rate and the real frequency decrease with increasing $L_{ez}$, and the mode propagates in the direction of the ion diamagnetic drift. Higher $f_c$ causes smaller (greater) growth rate for $L_{ez} > (<) L_{ez,c} \approx 0.27$, which is qualitatively consistent with Refs. 37, 44. The critical $L_{ez}$ for distinguishing the destabilization and stabilization effects is also in agreement with Eq. (10). Interestingly, isotopic effects on ITG instability are found to be weak for the typical parameters taken in this work, but $\hat{\gamma}_H > \hat{\gamma}_D > \hat{\gamma}_T$ is still satisfied as presented in Fig. 1(c). This is also qualitatively consistent with the numerical result of ITG modes in H, D, T plasmas with impurities reported in Ref. 53. The weak isotopic effects on ITG instability mainly due to the weak $A_i$ dependence of the effective ITG drive, $\overline{K}\tilde{f}_z = (1 - Zf_c)L_{ei}K_i + \frac{A_z}{A_i Z}f_c L_{ez}K_z$. Although impurity mass to ion mass ratio $A_z / A_i$ in the latter term is decreased from H to D, and to T, this term due to small impurity concentration is small as compared to the first term. Meanwhile, impurity influence on mode structure can be extracted from Eq. (5) with the simplified form $\delta\phi_k(x) \sim \exp(-i\frac{|S|}{\hat{\omega}}\frac{x^2}{2})$ under $F(\hat{\omega}) \approx 1$. For fixed $|S|$, the radial mode width proportional to $|\hat{\omega}|^{1/2}$ will be unambiguously influenced by changing impurity profile or impurity concentration as shown in Fig. 1(b). From Fig. 1(d), we can also observe that the isotopic effects slightly decrease the radial mode width in the presence of impurity because of the weak relationship $|\hat{\omega}_H| > |\hat{\omega}_D| > |\hat{\omega}_T|$.

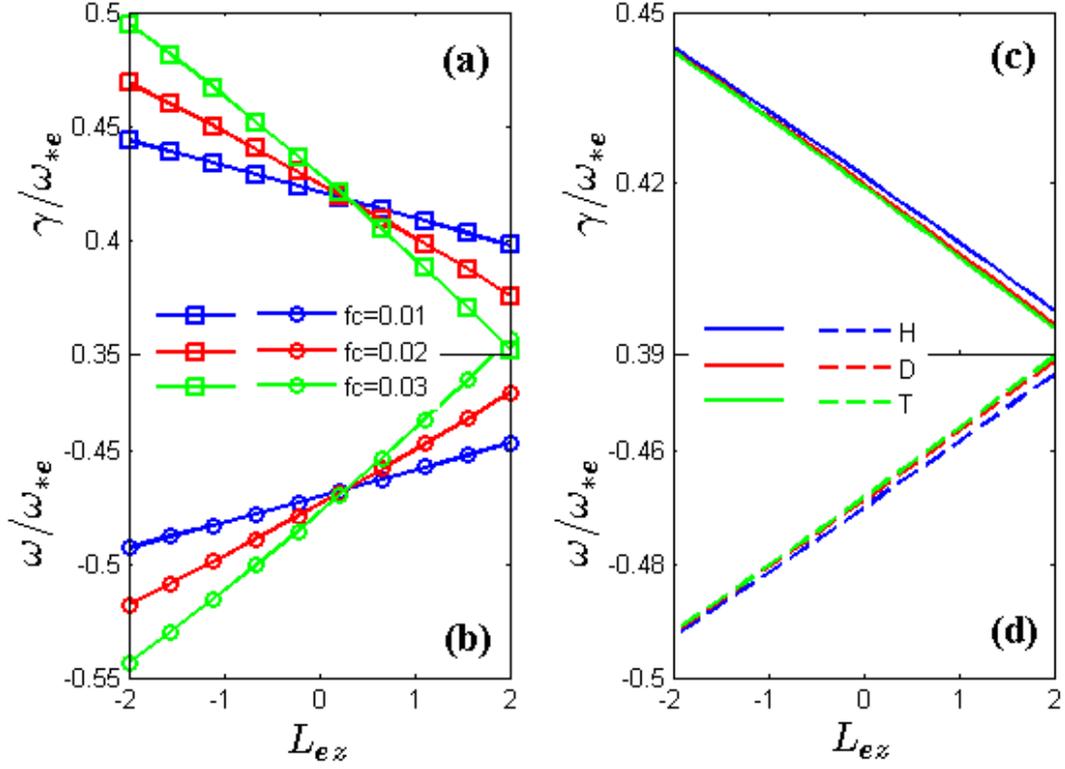

Fig. 1. Normalized growth rate and mode frequency as a function of $L_{ez}$ with different impurity concentration in Fig1. (a) and (b) and with different isotopic components of H, D, T in Fig1. (c) and (d).

### III. Isotopic dependence of impurity transport

QL turbulent impurity flux driven by radial fluctuating $E \times B$ velocity can be written as

$$\Gamma_z = \mathrm{Re}\langle \delta v_{E\times B,r}\delta n_z\rangle = -D_z \nabla n_{0z} + V_z n_{0z}, \tag{11}$$

with diffusivity

$$D_z = \sum_k \frac{|\gamma_k|}{4\omega_r^2}\left\{1 + \frac{1}{\tau_z}\frac{A_z}{A_i Z^2}\left\{\frac{\hat{\omega}_r}{2\hat{\gamma}_k}\frac{k_\parallel c_s}{\omega_r}\frac{1}{x}(2\ell+1) + \frac{3}{2}\left(\frac{A_i^2 Z^2}{A_z^2}-1\right)\frac{k_\parallel^2 c_s^2}{\omega_r^2} - b_{sy}\right\}\right\}k_y^2 \rho_s^2 c_s^2 |\phi_k|^2 \tag{12}$$

,

and convective velocity

$$V_z = \sum_k \frac{|\gamma_k|}{4\omega_r^2} \frac{A_z}{A_i Z} \left\{ \frac{1}{L_{n_e}} \hat{\omega}_r \left[ \frac{\hat{\omega}_r}{2\hat{\gamma}_k} \frac{k_\parallel c_s}{\omega_r} + \left( \frac{A_i^2 Z^2}{A_z^2} - 1 \right) \frac{k_\parallel^2 c_s^2}{\omega_r^2} \right] \right.$$

$$\left. + \frac{1}{L_{T_z}} \frac{1}{Z\tau_z} \left[ \frac{\hat{\omega}_r}{2\hat{\gamma}_k} \frac{k_\parallel c_s}{\omega_r} (2\ell+1) + \frac{3}{2} \left( \frac{A_i^2 Z^2}{A_z^2} - 1 \right) \frac{k_\parallel^2 c_s^2}{\omega_r^2} - b_{sy} \right] \right\} k_y^2 \rho_s^2 c_s^2 |\phi_k|^2$$

. (13)

Here, $b_{sy} = \frac{1}{2} k_y^2 \rho_s^2$, $\phi_k = \frac{e\delta\phi_k}{T_e}$. Likewise, the ion flux $\Gamma_i$ can be also calculated.

The radial ambipolar relationship of turbulent particle flux $\Gamma_i + Z\Gamma_z = 0$ can thus be easily verified for the adiabatic electrons. Many previous works on impurity transport are indirectly deducted from the ion flux through this ambipolar relationship. [46,59] On the right hand side of Eq. (13), the terms proportional to $1/L_{n_e}$ are denoted by $V_z^{\nabla n_e}$, and the terms proportional to $1/L_{T_z}$ are thermodiffusion convective velocity $V_z^{\nabla T}$. Here, we take orderings $k_\parallel^2 c_s^2 / \omega_r^2 \sim b_{sy} \sim \varepsilon$ with $\varepsilon$ being the small ordering parameter. Thus, all the terms related to parallel compression and $b_{sy}$ can be neglected. Then normalizing the flux to $n_{0z} c_s |\phi_k|^2$, the normalized transport coefficients ($\hat{D}_z = D_z/(R c_s |\phi_k|^2)$, $\hat{V}_z = V_z/(c_s |\phi_k|^2)$) are given as

$$\hat{D}_z = \sum_k \frac{|\hat{\gamma}_k|}{2\hat{\omega}_r^2} \frac{L_{ne}}{R} \left\{ 1 + \frac{1}{\tau_z} \frac{A_z}{A_i Z^2} \frac{1}{\hat{\gamma}_k} |S| \right\} k_y \rho_s, \quad (14)$$

$$\hat{V}_z = \sum_k \frac{1}{2\hat{\omega}_r^2} \frac{A_z}{A_i Z} \frac{L_{ne}}{R} \left\{ \frac{R}{L_{ne}} \hat{\omega}_r + \frac{R}{L_{T_z}} \frac{1}{Z\tau_z} \right\} \sqrt{|S|} k_y \rho_s, \quad (15)$$

where $\ell = 0$ is used. The normalized width of modes localized around the rational surface is taken as $\bar{x} = \sqrt{L_s/L_{n_e}}$. From Eq. (15), we can clearly see that the direction of convective velocity depends on the competition between inward $\hat{V}_z^{\nabla n_e}$ ($\hat{\omega}_r < 0$ for ITG) and outward $\hat{V}_z^{\nabla T}$. If $|\hat{\omega}_r| < \frac{1}{Z\tau_z} \frac{L_{n_e}}{L_{T_z}}$, the convective velocity $\hat{V}_z$ is outward, which is favorable for expulsion of impurities. Otherwise, $\hat{V}_z$ is inward, and may cause the accumulation of impurities in the core. Since turbulence amplitude itself can be a function of isotopes, the analysis on the normalized flux and transport coefficients for non-trace impurity is more useful for determining their signs than their absolute value. However, the analysis on the peaking factor for trace impurity may be justified due to the cancellation of amplitude dependence. In the following, the isotopic effects on impurity flux and transport coefficients with various parameters scan will be discussed.

**A. Non-trace impurity transport**

In this subsection, we take core peaked impurity profile, $L_{ez}=1$. The parametric dependence of isotopic effects on impurity transport with hollow impurity profile has the similar trends as the case with peaked impurity profile, so we only present the latter case. In Fig. 2, the normalized impurity flux and the corresponding transport coefficients are given as a function of $Z$ of Ar with different hydrogenic plasmas. The outward normalized impurity flux as well as $\hat{D}_z$ and $\hat{V}_z$ decreases and tends to saturate with the increase of $Z$. Particularly, $\hat{V}_z$ changes from outward to slightly inward with increasing $Z$. This is because the inward $\hat{V}_z^{\nabla n_e}$ ($\sim \frac{1}{Z}$) overcomes the outward $\hat{V}_z^{\nabla T}$ ($\sim \frac{1}{Z^2}$) with increasing $Z$ for fixed $A_z$. But, the outward diffusive part is larger than the inward convective part. Therefore, the flux is still outward which may have an ability of expelling impurity out. For lower $Z$ ($Z \leq 5$), changing from H to D, and to T plasmas, the isotopic effects reduce the outward impurity flux as shown in Fig. 2 (a). We can see that it is mainly because the outward convective velocity $\hat{V}_z$ is decreased by the isotopic effects as shown in Fig. 2 (b). This is also consistent with Eq. (15). The isotopic effects on flux and the corresponding transport coefficients are relatively weak for higher $Z$ ($Z>5$).

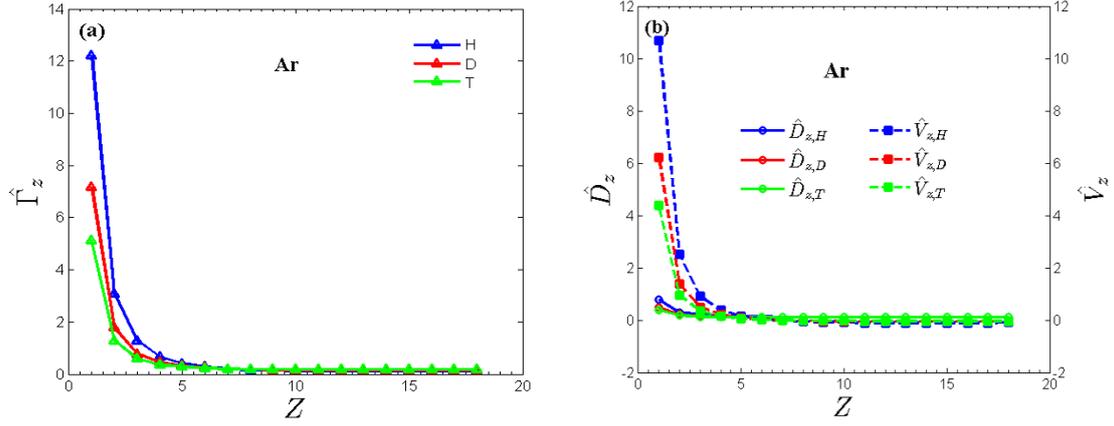

Fig. 2. Normalized impurity flux (a) and normalized transport coefficients (b) versus charge number $Z$ of Ar for H, D, T plasmas.

The mass number $A_z$ scan for isotopic effects on normalized impurity flux and the corresponding transport coefficients is presented in Fig. 3. Here, $A_z$ is the realistic mass number of fully ionized impurities from He to Ar. Impurity flux is always outward. The outward convective velocity $\hat{V}_z$ decreases with $A_z$, and changes to be inward, while the diffusivity $\hat{D}_z$ is almost unchanged with $A_z$ as shown in Fig. 3(b). These lead to the decrease of impurity flux with $A_z$ as shown in Fig. 3(a). Notably, the isotopic effects of impurity transport depend on impurity mass number $A_z$. The flux of lighter fully ionized impurities from He to C is reduced by isotopic effects. While, for the relatively heavier fully ionized impurities from N to Ar, the isotopic effects slightly enhance the flux. These are mainly because the outward (inward) convective velocity $\hat{V}_z$ for lighter (relatively heavier) impurities is decreased by isotopic effects as shown in Fig. 3(b). In one word, the isotopic effects are favorable (unfavorable) for relatively heavier (lighter) impurity transport.

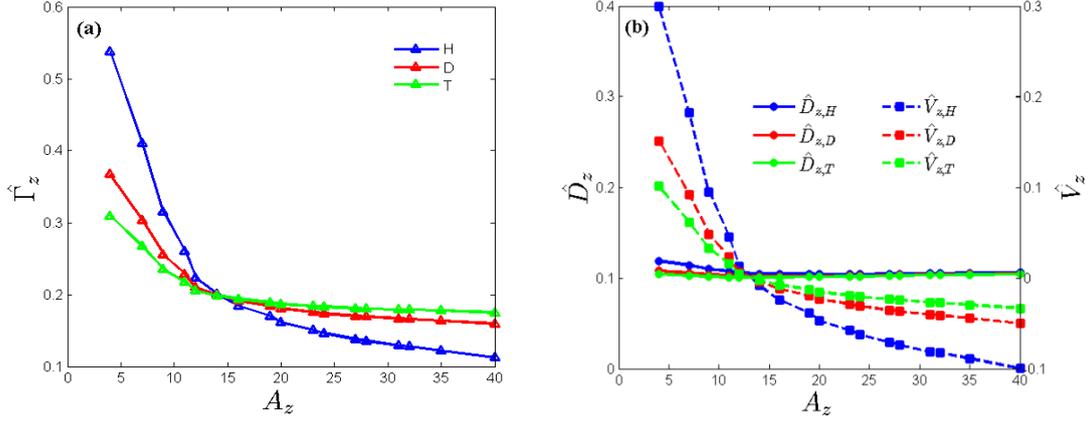

Fig. 3. Normalized impurity flux (a) and normalized transport coefficients (b) versus realistic mass number of fully ionized impurities from He to Ar with H, D and T plasmas.

In Fig. 4, we show the magnetic shear parameter dependence $|S|$ on impurity flux and the corresponding transport coefficients for H, D and T plasmas. The outward flux as well as $\hat{D}_z$ and outward $\hat{V}_z$ for H, D and T are all decreasing with $|S|$. The convective velocity $\hat{V}_z$ even changes from outward to inward with increasing $|S|$. We can see that the isotopic effects on impurity transport also depend on magnetic shear. For weak magnetic shear ($|S|<0.15$), the outward flux is reduced by isotopic effects, but enhanced by isotopic effects for strong magnetic shear ($|S|\geq 0.15$) as shown in Fig. 4(a). These are mainly because the isotopic effects reduce the absolute value of $\hat{V}_z$ for the whole regime of $|S|$. In other words, the isotopic effects can enhance impurity transport if the absolute value of magnetic shear is strong.

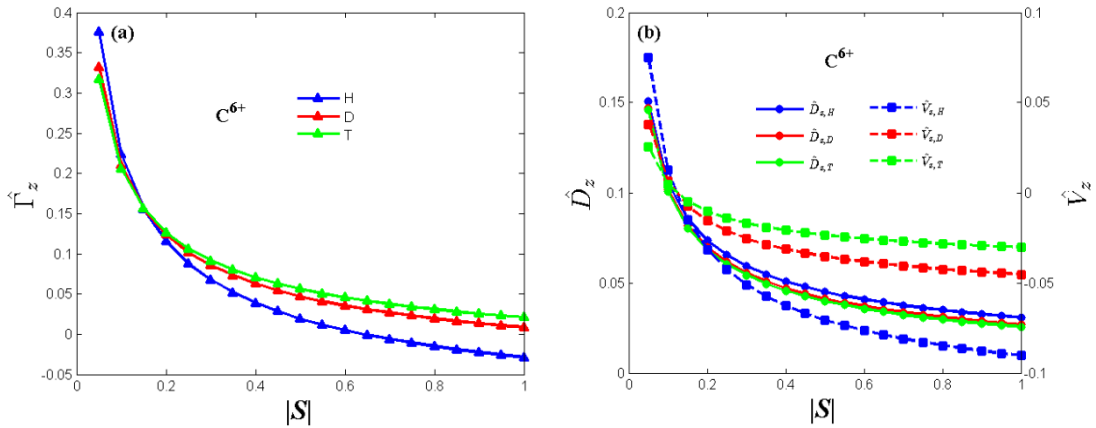

Fig. 4. Normalized impurity flux (a) and normalized transport coefficients (b) versus absolute value of magnetic shear parameter $|S|$ for H, D, T plasmas.

In Fig. 5, the impurity flux and the normalized transport coefficients versus electron-to-impurity temperature ratio $\tau_z$ are illustrated. Here, the impurity is high temperature $He^{2+}$ from D-T reaction. As we know, the alpha particles from D-T reaction dominantly heat the electrons, [60] and then exchange energy to ions by collisions. As the energy exchange time is typically shorter than the slowing down time of alpha particles, so we assume $\tau_i = 1$, $\tau_z \leq 1$ for $He^{2+}$, and take the tolerance concentration as 10% in ITER according to Ref. 3. Here, we simply assumed that the temperature gradient of $He^{2+}$ equals to that of main ions, then we have $\eta_z = \eta_i \frac{\tau_z}{\tau_i} \left( \frac{1}{L_{ez}} - Zf_c \right) / (1 - Zf_c)$. The pair of isotopic fueling ratios $f_D = n_{0D} / (n_{0D} + n_{0T})$ and $f_T = n_{0T} / (n_{0D} + n_{0T})$ are set as 25%+75%, 50%+50% and 75%+25%, respectively, and $A_i$ in Eqs. (8), (9), (14) and (15) for D-T plasmas is replaced by effective ion mass number $A_{i,eff} = 2f_D + 3f_T$. It can be clearly seen that the flux is decreased with increasing $\tau_z$ (decreasing $T_z$) as shown in Fig. 5 (a). It indicates that it is easier for transporting higher temperature $He^{2+}$ out. This might be a bad news for ITER. In Fig. 5 (b), with increasing $\tau_z$, the diffusivity $\hat{D}_z$ apparently decreases, while the outward convective velocity $\hat{V}_z$ increases to saturation. The decrease of $\hat{D}_z$ dominates the increase of $\hat{V}_z$. For high temperature $He^{2+}$ in Fig. 5 (b), the isotopic effects prominently reduce the outward convective velocity $\hat{V}_z$, but show a weak influence on the diffusivity $\hat{D}_z$. Therefore, the outward flux is reduced by isotopic effects as shown in Fig. 5 (a).

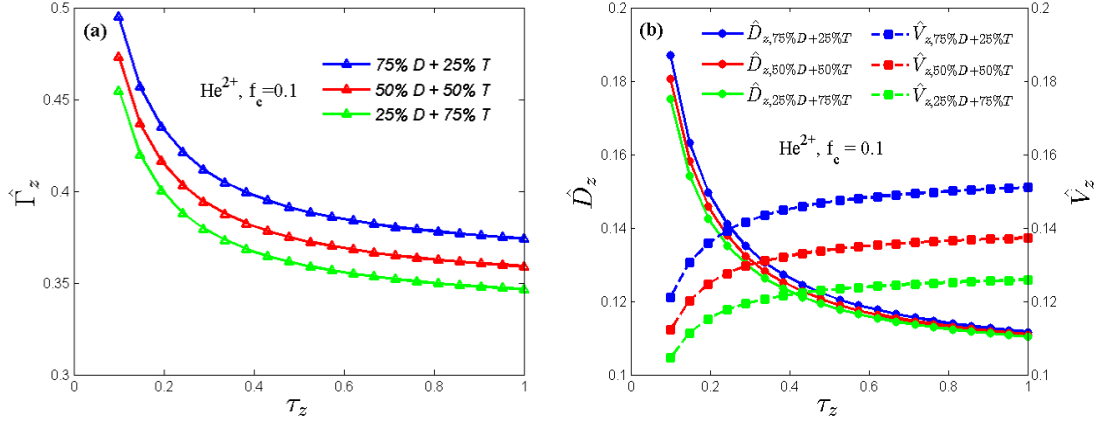

Fig. 5. Normalized impurity flux (a) and normalized transport coefficients (b) versus electron-to-impurity temperature ratio $\tau_z$ for $He^{2+}$ impurity from D-T reaction. The concentration for high temperature $He^{2+}$ is taken as $f_c = 0.1$.

**B. Trace impurity transport**

As we know, W will be used in ITER divertor, and could be also considered in the future reactor. The tolerance of W in ITER and present ITER-like wall (ILW) devices is very limited.[3] Here, we set W concentration as $f_c = 10^{-4}$, and W can thus be treated as trace impurity. The isotopic effects on PF, $-RV_z/D_z$ with scans of $Z$ and $|S|$ are shown in Fig. 6. The negative (positive) PF corresponds to outward (inward) $V_z$. In Fig. 6 (a), the absolute value of negative PF is reduced, and then changes sign to positive with the increase of $Z$. This may suggest accumulation of higher ionized W. But, the good news is that the accumulation of higher ionized W can be reduced by isotopic effects. However, it is noted that the transport of high-$Z$ impurities are higher with higher effective charge number and higher effective mass number in JET.[51] In Fig. 6(b), we show that the accumulation (positive PF) of $W^{45+}$ is enhanced by increasing $|S|$. The isotopic effects can reduce the PF of $W^{45+}$, i.e., lead to reduction of accumulation. Moreover, the isotopic effects get stronger with the increase of $|S|$.

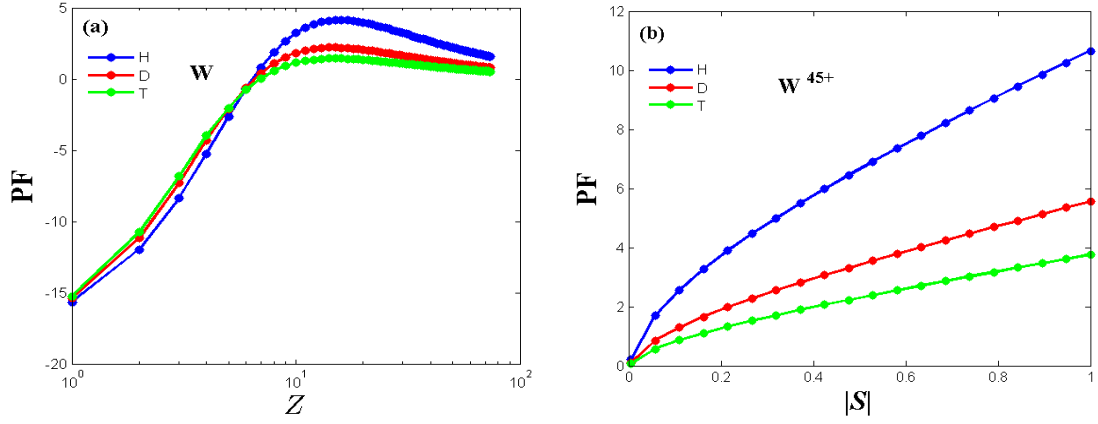

Fig. 6. Impurity peaking factor $-RV_z/D_z$ versus $Z$ (a) and $|S|$ (b) for H, D, T plasmas.

## IV. Summary and discussions

In the present work, isotopic effects on non-trace and trace impurity transport driven by electrostatic ITG turbulence in sheared slab geometry are investigated. QL impurity flux divided into the summation of diffusion and convection is calculated. Then the isotopic effects on non-trace impurity flux and the corresponding transport coefficients are analyzed. For trace impurity, the isotopic effects on PF are also discussed. The principal results of this work are as follows.

(1) Isotopic effects on ITG mode are found to be weak with the presence of impurity. In particular, the critical impurity density gradient for discriminating destabilization and stabilization effects of impurity is achieved as given in Eq. (10).

(2) It is found that the isotopic effects are unfavorable (favorable) for transporting impurity with relatively lower (higher) $Z$, lighter (heavier) mass and weaker (stronger) magnetic shear. This is mainly because the outward (inward) velocity $\hat{V}_z$ are decreased by isotopic effects, while isotopic effects on diffusivity is relatively weak.

(3) It is easier for expelling higher temperature $He^{2+}$ from D-T reaction, which might be a bad news for ITER. Moreover, isotopic effects reduce the outward flux of high temperature $He^{2+}$, which is also mainly because the outward convective velocity $\hat{V}_z$ is reduced by isotopic effects.

(4) For trace W, the decreases of $Z$ and $|S|$ reverse the $V_z$ from inward to outward, which is favorable for easing the accumulation degree in the core. The isotopic effects also apparently reduce the accumulation level of high-$Z$ impurities such as $W^{45+}$, and these effects get stronger with the increase of $|S|$.

The major efforts of the present work are aimed at investigating the isotopic effects on impurity transport driven by ITG turbulence in slab geometry. More carefully study in realistic tokamak geometry and extension to electromagnetic micro-turbulence are also worthwhile. Not only the mode structure and growth rate can be modified by toroidal effects, but also the turbulent flux changes. Especially, the isotopic effects on turbulent equi-partition (TEP) pinch seems an interesting topic in tokamak plasmas. A further step could include the nonlinear coupling of ITG and zonal flows. With powerful electron heating and alpha particle heating, it is also interesting and necessary to study the isotopic effects on impurity transport driven by other candidates for anomalous transport, such as TEM and electron temperature gradient (ETG) turbulence. Moreover, global gyro-kinetic or gyro-fluid simulation about the isotopic effects on impurity transport is also an open question. All these possible subjects are left for the future.

**ACKNOWLEDGMENTS**

We are grateful to P. H. Diamond, Q. Yu, X. Garbet, J. Q. Dong and the participants in the Festival of Theory, Aix en Provence 2015 and 43rd European Physical Society Conference on Plasma Physics, Leuven, Belgium 2016 for fruitful discussions. This work was supported by the NSFC Grant Nos. 11675059 and 11305071, the MOST of China under Contract No. 2013GB112002.